\documentclass[aps,prb,twocolumn,amsmath,amssymb]{revtex4}

\usepackage{graphicx}
\usepackage{bm}
\usepackage{bbold}
\usepackage{hhline}

\def\bk{ \bm{k} }
\def\bmd{ \bm{d} }

\begin{document}
\title{Exotic interband pairing in multiband superconductors}

\author{K. V. Samokhin\footnote{E-mail: kirill.samokhin@brocku.ca}}
\affiliation{Department of Physics, Brock University, St. Catharines, Ontario L2S 3A1, Canada}


\begin{abstract}
Contrary to the usual assumption, the electron Bloch states in crystals with spin-orbit coupling do not always transform under symmetry operations in the same way as the pure
spin-$1/2$ states. This has profound consequences for the symmetry properties and nodal structure of superconductors, especially for the interband gap functions. 
Focusing on tetragonal superconductors, we show that the interband pairing in the conventional ($s$-wave) channel can have features which are traditionally associated with unconventional pairing, 
such as triplet components and odd parity, and can produce line nodes in the excitation energy gap.
In the $d$-wave case, the interband pairing, which can also be odd in momentum and have a triplet component, changes the positions and topology of the nodal lines.
\end{abstract}

\maketitle

\section{Introduction}
\label{sec: Intro}

The symmetry-based phenomenological approach has proved to be very useful in the studies of fermionic superfuilds and superconductors.\cite{VG85,SU-review,TheBook} This approach allows one
to determine the stable states and gap structures even if the pairing mechanism is not fully understood. 
The idea at the heart of the symmetry approach is that the electron Bloch states in the presence of spin-orbit coupling transform under the crystal point group operations and time reversal (TR) 
in the same way as the pure spin-$1/2$ states,\cite{And84,UR85} leading to the relatively simple transformation rules for the superconducting gap functions. 

The electron bands $\xi_n(\bk)=\xi_n(-\bk)$ in a nonmagnetic centrosymmetric crystal are twofold degenerate at each wave vector $\bk$ due to the conjugation symmetry ${\cal C}=KI$ (Ref. \onlinecite{Kittel-book}), 
which combines TR operation $K$ with space inversion $I$ and leaves $\bk$ invariant. The bands are labelled by $n$, while an additional index $s=1,2$, called the conjugacy index, distinguishes two orthogonal 
states $|\bk,n,1\rangle$ and $|\bk,n,2\rangle\equiv{\cal C}|\bk,n,1\rangle$ within the same band. There is still some freedom is choosing the relative ``orientations'' of the Bloch bases at different $\bk$ points.
The usual assumption, formalized by the Ueda-Rice prescription,\cite{UR85} is that the bases at $\bk$ and $g\bk$, where $g$ is an element of the crystal point group $\mathbb{G}$,
are related by the same spin rotation matrix $\hat D^{(1/2)}(g)$ as the pure spin states, thus justifying the name ``pseudospin'' for the conjugacy index $s$. 
Then, the pseudospin-singlet (pseudospin-triplet) superconducting gaps respond to the point group operations like scalar (pseudovector) functions.

Recently, however, there has been mounting evidence that the textbook classification might fail in superconductors with a complicated electronic structure, e.g., in multiorbital systems.\cite{multiorbital-SC} 
Also, the ``$j=3/2$'' pairing proposed for half-Heusler compounds\cite{j-3-2-pairing} is clearly outside the scope of the pseudospin-based approach. In this paper, we examine the validity of the 
pseudospin picture and show how a non-pseudospin character of the electron Bloch states changes the pairing symmetry and the gap nodal structure, focusing especially on the unusual features and effects of the interband pairing.
The intraband pairing in non-pseudospin bands was studied in Ref. \onlinecite{Sam19-PRB}.
We use the band representation, motivated by its importance for any Bardeen-Cooper-Schrieffer-like theory of superconductivity, in which fermionic quasiparticles exist and experience an attractive interaction 
only in the vicinity of the Fermi surfaces. 

The paper is organized as follows. In Sec. \ref{sec: Gap symmetry}, we construct the Bloch bases in momentum space in the way which is consistent with the point-group symmetry requirements and also derive the general
symmetry constraints on the superconducting gap functions. In Sec. \ref{sec: Interband}, we study the interband gap symmetry in tetragonal superconductors. The effects of the interband pairing on the 
Bogoliubov excitation spectrum in $s$-wave and $d$-wave superconductors are discussed in Sec. \ref{sec: Gap nodes}. Our findings are summarized in Sec. \ref{sec: Conclusions}.
Throughout the paper we use the units in which $\hbar=1$, neglecting, in particular, the difference between the quasiparticle wave vector and momentum.

\section{Bloch bases and gap symmetry}
\label{sec: Gap symmetry}

We start with the general mean-field pairing Hamiltonian $\hat H=\sum_{\bk ns}\xi_n(\bk)c^\dagger_{\bk ns}c_{\bk ns}+\hat H_{sc}$, where
\begin{eqnarray}
\label{H-mean-field-general}
  \hat H_{sc} = \frac{1}{2}\sum_{\bk,nn',ss'}\left[\Delta_{nn',ss'}(\bk)c^\dagger_{\bk ns}\tilde c^\dagger_{\bk n's'}+\mathrm{H.c.}\right],
\end{eqnarray}
with $\tilde c^\dagger_{\bk ns}=Kc^\dagger_{\bk ns}K^{-1}$.
The presence of the antiunitary TR operation in $\hat H_{sc}$ is crucial for the proper definition of the gap functions $\hat\Delta_{nn'}(\bk)$ (Ref. \onlinecite{BlountSam}). 
The $2\times 2$ matrices $\hat\Delta_{nn}$ describe the intraband pairing in the $n$th band, while $\hat\Delta_{nn'}$ with $n\neq n'$ describe the interband pairing. The latter may become important in strongly-coupled 
superconductors or when the pairing in a multiorbital system is translated into the band representation. 

According to the Landau theory of phase transitions, the gap functions transform according to a single-valued irreducible representation (irrep) $\gamma$ of the point group $\mathbb{G}$. 
In order to find their momentum dependence, in particular, the type and location of the gap nodes, one has to know how the gaps are affected by the 
crystal symmetry operations, i.e., how $\hat\Delta_{nn'}(\bk)$ is related to $\hat\Delta_{nn'}(g\bk)$, where $g\in\mathbb{G}$, which in turn depends on the transformation properties of the single-particle Bloch states.

The band symmetry at the $\Gamma$ point is described by the magnetic point group ${\cal G}=\mathbb{G}+{\cal C}\mathbb{G}$ and the Bloch states there
form the basis of an irreducible double-valued corepresentation (corep) of ${\cal G}$, see, e.g., Ref. \onlinecite{BC-book}. The conjugation operation is antiunitary, hence the group is ``magnetic'', and satisfies ${\cal C}^2=-1$ 
when acting on spinor wave functions. In a centrosymmetric crystal, the coreps of ${\cal G}$ are either inversion-even ($\Gamma^+$) or inversion-odd ($\Gamma^-$). 
An important observation is that the double-valued coreps of the crystallographic magnetic point groups are almost all two-dimensional (2D). 
There are just two exceptions: $(\Gamma_6^\pm,\Gamma_7^\pm)$ for $\mathbb{G}=\mathbf{T}_{h}$ and $\Gamma_8^\pm$ (``$j=3/2$'') for $\mathbb{G}=\mathbf{O}_{h}$, which are four-dimensional and 
will not be considered here. 

We assume that the Bloch states $|\bm{0},n,1\rangle$ and $|\bm{0},n,2\rangle$ transform according to a 2D double-valued corep of ${\cal G}$ described by $2\times 2$ matrices $\hat{\cal D}_n(g)$. 
To construct the Bloch bases at $\bk\neq\bm{0}$ satisfying all symmetry and compatibility requirements, we start
with some wave vector $\bk$ and apply a point-group element $g$ to transform $\bk$ into $g\bk$. 
The state $g|\bk,n,s\rangle$ belongs to the wave vector $g\bk$ and can be represented as $g|\bk,n,s\rangle=\sum_{s'}|g\bk,n,s'\rangle U_{n,s's}(\bk;g)$,
where the expansion coefficients form a unitary matrix. Since in the case under consideration the Bloch wave functions and the band dispersions are analytic functions of $\bk$,
one can choose the $U$ matrix to be $\bk$ independent, at least in the vicinity of the $\Gamma$ point. Therefore, 
$\hat U_n(g)=\hat{\cal D}_n(g)$ and the Bloch basis at $g\bk$ is defined by the following relation:\cite{Sam19-PRB}
\begin{equation}
\label{Bloch-bases}
  g|\bk,n,s\rangle=\sum_{s'}|g\bk,n,s'\rangle {\cal D}_{n,s's}(g),\quad g\in\mathbb{G}. 
\end{equation}
In particular, $I|\bk,n,s\rangle=p_n|-\bk,n,s\rangle$, where $p_n=\pm$ is the parity of the band. 

If the $\Gamma$-point corep is equivalent to the spin-$1/2$ corep, then the band is
called ``pseudospin band'' and $\hat{\cal D}_n(g)=\hat D^{(1/2)}(R)$ ($g$ is either a proper rotation $R$ or an improper rotation $IR$), corresponding to the Ueda-Rice convention.\cite{UR85}
In general, the Bloch states at the $\Gamma$ point transform according to a corep which is not equivalent to the spin-$1/2$ corep, i.e., $\hat{\cal D}_n(g)\neq\hat D^{(1/2)}(R)$ and we have a ``non-pseudospin'' band. 
The effects of non-pseudospin character of the bands can be seen already in the normal state, e.g., in the form of the antisymmetric spin-orbit coupling in crystals without an inversion center.\cite{unusual-ASOC}

From Eq. (\ref{Bloch-bases}), the point-group transformation rules for the electron creation operators take the form
$gc^\dagger_{\bk ns}g^{-1}=\sum_{s'}c^\dagger_{g\bk,ns'}{\cal D}_{n,s's}(g)$. Applying this to the pairing Hamiltonian (\ref{H-mean-field-general}), we find that the gap functions transform in the following way:
\begin{equation}
\label{Delta-transform-g}
  g:\ \hat\Delta_{nn'}(\bk)\to\hat{\cal D}_n(g)\hat\Delta_{nn'}(g^{-1}\bk)\hat{\cal D}^\dagger_{n'}(g),\quad g\in\mathbb{G}.
\end{equation}
Since $\tilde c^\dagger_{\bk n1}=p_nc^\dagger_{-\bk,n2}$ and $\tilde c^\dagger_{\bk n2}=-p_nc^\dagger_{-\bk,n1}$, we obtain the response to TR:
\begin{equation}
\label{Delta-transform-K}
  K:\ \hat\Delta_{nn'}(\bk)\to\hat\Delta^\dagger_{n'n}(\bk),
\end{equation}
and also a constraint on the gap functions which follows from the anticommutation of the fermionic operators:
\begin{equation}
\label{Delta-constraint-anticommutation}
  \hat\Delta_{nn'}(\bk)=p_np_{n'}\hat\sigma_2\hat\Delta^\top_{n'n}(-\bk)\hat\sigma_2,
\end{equation} 
where $\hat{\bm\sigma}$ are the Pauli matrices in the conjugacy space. Note that the anticommutation constraint does not select the parity of the interband pairing, see more on that below. 
According to Eq. (\ref{Delta-transform-g}), the gap transformation properties are nonuniversal, in the sense that they essentially depend on the symmetries of the bands $n$ and $n'$ involved in the pairing. 

One can see that even the intraband pairing may be rather unusual.\cite{Sam19-PRB} Let us drop the band index and represent the gap function as a sum of the singlet and triplet components: 
$\hat\Delta(\bk)=\psi(\bk)\hat\sigma_0+\bmd(\bk)\hat{\bm{\sigma}}$ ($\hat\sigma_0$ is the $2\times 2$ unit matrix in the conjugacy space). 
The parity follows from the constraint (\ref{Delta-constraint-anticommutation}): $\psi(-\bk)=\psi(\bk)$ and $\bmd(-\bk)=-\bmd(\bk)$, whereas Eq. (\ref{Delta-transform-g}) yields $\psi(\bk)\to\psi(g^{-1}\bk)$ 
and $\bmd(\bk)\to{\cal R}(g)\bmd(g^{-1}\bk)$,
where $\hat{\cal R}$ is the $3\times 3$ orthogonal matrix defined by $\hat{\cal D}^\dagger(g)\hat{\sigma}_i\hat{\cal D}(g)=\sum_{j=1}^3{\cal R}_{ij}(g)\hat{\sigma}_j$.
Thus, $\psi$ transforms as a complex scalar regardless of the band symmetry at the $\Gamma$ point, and the usual classification of the singlet superconducting states\cite{VG85,SU-review,TheBook} is applicable. 
In contrast, the transformation properties of the triplet gap depend on the band symmetry. Namely, if the $\Gamma$-point corep is such that $\hat{\cal R}\neq\hat R$, where 
$\hat R\equiv\hat D^{(1)}(R)$ is the spin-$1$ rotation matrix, then $\bmd$ does not transform as a pseudovector, which strongly affects its momentum dependence. 
As shown in Ref. \onlinecite{Sam19-PRB}, this happens in certain bands in trigonal and hexagonal superconductors.

\subsection{Basis functions}
\label{sec: basis functions}

Different pairing channels correspond to different single-valued irreps of the point group $\mathbb{G}$ of the crystal. In order to determine the momentum dependence of the pairing involving quasiparticles 
from the bands $n$ and $n'$, we observe that the gap function corresponding to a $d$-dimensional irrep $\gamma$ can be represented in the form
\begin{equation}
\label{irrep-expansion}
  \hat\Delta_{nn'}(\bk)=\sum_{a=1}^d\eta_{nn'}^{a}\hat\phi_{nn'}^{a}(\bk),
\end{equation}
where $\eta_{nn'}^{a}$ are the superconducting order parameter components, which are found by minimizing the free energy of the superconductor, and $\hat\phi_{nn'}^{a}$ are the $2\times 2$ matrix basis functions of $\gamma$. 
According to Eq. (\ref{Delta-transform-g}), the latter transform under the point group operations as follows:
\begin{eqnarray}
\label{phi-transform-g}
  g:\hat\phi_{nn'}^{a}(\bk)\to\hat{\cal D}_n(g)\hat\phi_{nn'}^{a}(g^{-1}\bk)\hat{\cal D}^\dagger_{n'}(g) \nonumber \\
  =\sum_{b=1}^d\hat\phi_{nn'}^{b}(\bk)D_{\gamma,ba}(g),
\end{eqnarray}
where $\hat{\cal D}_n$ is the $\Gamma$-point corep matrix and $\hat D_\gamma$ is the irrep matrix. In particular, setting $g=I$ in Eq. (\ref{phi-transform-g}), we obtain:
\begin{equation}
\label{phi-transform-I}
  p_np_{n'}\hat\phi_{nn'}^{a}(-\bk)=P_\gamma\hat\phi_{nn'}^{a}(\bk),
\end{equation}
where $P_\gamma$ is the parity of $\gamma$. It follows from the anticommutation constraint (\ref{Delta-constraint-anticommutation}) that
$\hat\phi_{nn'}^{a}(\bk)=p_np_{n'}\hat\sigma_2\hat\phi^{a,\top}_{n'n}(-\bk)\hat\sigma_2$, therefore, $\eta_{nn'}^{a}=\eta_{n'n}^{a}$ and, using Eq. (\ref{phi-transform-I}), we have
\begin{equation}
\label{phi-constraint-AC-I}
  \hat\phi_{nn'}^{a}(\bk)=P_\gamma\hat\sigma_2\hat\phi_{n'n}^{a,\top}(\bk)\hat\sigma_2.
\end{equation}
Regarding the response to TR, see Eq. (\ref{Delta-transform-K}), the basis functions can be chosen to satisfy 
\begin{equation}
\label{phi-constraint-TR}
  \hat\phi_{nn'}^{a}(\bk)=\hat\phi_{n'n}^{a,\dagger}(\bk),
\end{equation}
which means that the action of TR on the order parameter components is equivalent to complex conjugation.
In the next section, we apply the general expressions derived above to one-dimensional (1D) pairing channels in a tetragonal superconductor.

\section{Interband pairing}
\label{sec: Interband}

It is evident from Eq. (\ref{Delta-transform-g}) that there exists a multitude of possibilities for the interband pairing symmetry, depending on the 
$\Gamma$-point coreps in the bands $n$ and $n'$. We use as an example a tetragonal superconductor with $\mathbb{G}=\mathbf{D}_{4h}$, which is relevant for many popular materials, 
from the high-$T_c$ cuprates and iron pnictides to Sr$_2$RuO$_4$. Among the four double-valued coreps $\Gamma_6^\pm$ and $\Gamma_7^\pm$ of $\mathbf{D}_{4h}$, only $\Gamma_6^+$ is equivalent 
to the spin-$1/2$ corep,\cite{BC-book} while the other three are non-pseudospin ones. There are ten possible two-band combinations: $(n,n')=(\Gamma_6^+,\Gamma_6^+)$, $(\Gamma_6^+,\Gamma_6^-)$, $(\Gamma_6^+,\Gamma_7^+)$, etc. 

Regarding the pairing channel $\gamma$, the group $\mathbf{D}_{4h}$ has ten single-valued irreps of either parity, eight 1D and two 2D. 
We consider just two pairing channels: the conventional, or ``$s$-wave'', pairing corresponding to the identity irrep $A_{1g}$, and the unconventional ``$d_{x^2-y^2}$-wave'' pairing 
corresponding to the irrep $B_{1g}$ (Ref. \onlinecite{irreps}). Both irreps are 1D, so that Eq. (\ref{irrep-expansion}) takes the form 
\begin{equation}
\label{Delta-1D}
  \hat\Delta_{nn'}(\bk)=\eta_{nn'}\hat\phi_{nn'}(\bk), 
\end{equation}
where $\eta_{nn'}=\eta_{n'n}$ are the order parameter components. If there are $N$ superconducting bands, then the total number of independent components, intraband and interband, is equal to $N(N+1)/2$. 
According to Eq. (\ref{phi-transform-g}), the basis functions satisfy the following equation: 
\begin{equation}
\label{phi-transform-g-1D}
  \hat{\cal D}_n(g)\hat\phi_{nn'}(g^{-1}\bk)\hat{\cal D}^\dagger_{n'}(g)=\chi_\gamma(g)\hat\phi_{nn'}(\bk), 
\end{equation}
where $\chi_\gamma(g)$ is the character of $g$ in the 1D irrep $\gamma$. 

It is convenient to introduce the ``singlet-triplet'' decomposition of the basis functions: 
$$
\hat\phi_{nn'}(\bk)=\alpha_{nn'}(\bk)\hat\sigma_0+i\bm{\beta}_{nn'}(\bk)\hat{\bm{\sigma}}. 
$$
Since both the $s$-wave and $d$-wave irreps are even ($P_\gamma=1$), we obtain from Eqs. (\ref{phi-transform-I}), (\ref{phi-constraint-AC-I}), and (\ref{phi-constraint-TR}) that 
$\alpha$ and $\bm{\beta}$ are real and satisfy 
\begin{equation}
\label{ab-parities}
  \left.\begin{array}{l}
  \alpha_{nn'}(\bk)=\alpha_{n'n}(\bk)=p_np_{n'}\alpha_{nn'}(-\bk),\\ \\
  \bm{\beta}_{nn'}(\bk)=-\bm{\beta}_{n'n}(\bk)=p_np_{n'}\bm{\beta}_{nn'}(-\bk).
  \end{array}\right.
\end{equation}
Setting $n=n'$, we see that the intraband triplet components vanish, while the singlet components are even in $\bk$ and satisfy
$\alpha_{nn}(g^{-1}\bk)=\chi_\gamma(g)\alpha_{nn}(\bk)$. Therefore, the standard symmetry analysis\cite{SU-review,TheBook} is applicable for the intraband gaps. In contrast, the interband pairing structure can be 
considerably richer.

Focusing on just one pair of bands $n,n'=1,2$, we introduce the notation $\eta_{nn}=\eta_n$, $\alpha_{nn}=\alpha_n$, $\eta_{12}=\eta_{21}=\tilde\eta$, 
$\alpha_{12}=\alpha_{21}=\tilde\alpha$, and $\bm{\beta}_{12}=-\bm{\beta}_{21}=\tilde{\bm{\beta}}$, where $\alpha_1$, $\alpha_2$, $\tilde\alpha$, and $\tilde{\bm{\beta}}$ are all real functions of $\bk$. Then, the gap functions take the following form:
\begin{eqnarray}
\label{two-band-gaps}
        & \hat\Delta_{11}(\bk)=\eta_1\alpha_1(\bk)\hat\sigma_0,\quad\hat\Delta_{22}(\bk)=\eta_2\alpha_2(\bk)\hat\sigma_0, \nonumber \\
        & \hat\Delta_{12}(\bk)=\tilde\eta[\tilde\alpha(\bk)\hat\sigma_0+i\tilde{\bm{\beta}}(\bk)\hat{\bm{\sigma}}], \\
        & \hat\Delta_{21}(\bk)=\tilde\eta[\tilde\alpha(\bk)\hat\sigma_0-i\tilde{\bm{\beta}}(\bk)\hat{\bm{\sigma}}].\nonumber
\end{eqnarray}
Stable superconducting states correspond to the minima of the Ginzburg-Landau free energy, which is a functional of the order parameter components $\eta_1$, $\eta_2$, and $\tilde\eta$. 
Phenomenologically, it has the same form as in the three-band Ginzburg-Landau model, which has both TR invariant and TR symmetry-breaking stable states.\cite{ST10} 
We consider the general case, in which all three components are nonzero.

In the $s$-wave case, the intraband gaps are invariant under all $g$, e.g., $\alpha_n(\bk)\propto k_x^2+k_y^2+ak_z^2$ (we do not bother to normalize the basis functions), while
the interband basis functions for all possible pairs of bands are shown in Table \ref{table: phis-Gamma_1}.
Note that the interband pairing in the $s$-wave channel has features that are traditionally associated with unconventional pairing, such as symmetry-imposed zeros and triplet components, and it can be either even 
or odd in $\bk$, depending on the relative parity of the bands.
In the $d$-wave case, we have $\alpha_{n}(\bk)\propto k_x^2-k_y^2$, whereas the interband basis functions are listed in Table \ref{table: phis-Gamma_3}.

\begin{table*}
\caption{Momentum dependence of the interband pairing in an $s$-wave tetragonal superconductor, $a$ is a real constant. First column: the $\Gamma$-point coreps of the bands participating in the pairing.}
\begin{tabular}{|lll|}
    \hline
    $(n,n')$  & $\tilde\alpha(\bk)$ & $\tilde{\bm{\beta}}(\bk)$ \\ \hline
    $(\Gamma_6^\pm,\Gamma_6^\pm)$, $(\Gamma_7^\pm,\Gamma_7^\pm)$\hspace*{10mm} & $k_x^2+k_y^2+ak_z^2$  & $[k_yk_z,-k_xk_z,a(k_x^2-k_y^2)k_xk_y]$ \\ \hline
    $(\Gamma_6^\pm,\Gamma_6^\mp)$, $(\Gamma_7^\pm,\Gamma_7^\mp)$  &  $(k_x^2-k_y^2)k_xk_yk_z$\hspace*{10mm}   &  $(k_x,k_y,ak_z)$ \\ \hline
    $(\Gamma_6^\pm,\Gamma_7^\pm)$  &  $k_x^2-k_y^2$  &  $(k_yk_z,k_xk_z,ak_xk_y)$ \\ \hline
    $(\Gamma_6^\pm,\Gamma_7^\mp)$  &  $k_xk_yk_z$  &  $[k_x,-k_y,a(k_x^2-k_y^2)k_z]$ \\ \hline
\end{tabular}
\label{table: phis-Gamma_1}
\end{table*}

\begin{table*}
\caption{Momentum dependence of the interband pairing in a $d_{x^2-y^2}$-wave tetragonal superconductor, $a$ is a real constant. First column: the $\Gamma$-point coreps of the bands participating in the pairing.}
\begin{tabular}{|lll|}
    \hline
    $(n,n')$  & $\tilde\alpha(\bk)$ & $\tilde{\bm{\beta}}(\bk)$ \\ \hline
    $(\Gamma_6^\pm,\Gamma_6^\pm)$, $(\Gamma_7^\pm,\Gamma_7^\pm)$\hspace*{10mm}  &  $k_x^2-k_y^2$  &  $(k_yk_z,k_xk_z,ak_xk_y)$ \\ \hline
    $(\Gamma_6^\pm,\Gamma_6^\mp)$, $(\Gamma_7^\pm,\Gamma_7^\mp)$  &  $k_xk_yk_z$  &   $[k_x,-k_y,a(k_x^2-k_y^2)k_z]$ \\ \hline
    $(\Gamma_6^\pm,\Gamma_7^\pm)$  & $k_x^2+k_y^2+ak_z^2$   &  $[k_yk_z,-k_xk_z,a(k_x^2-k_y^2)k_xk_y]$ \\ \hline
    $(\Gamma_6^\pm,\Gamma_7^\mp)$  &  $(k_x^2-k_y^2)k_xk_yk_z$\hspace*{10mm}   & $(k_x,k_y,ak_z)$ \\ \hline
\end{tabular}
\label{table: phis-Gamma_3}
\end{table*}

As an example of the calculation of the interband gap functions, let us consider the $s$-wave pairing channel for the bands corresponding to the coreps $\Gamma_6$ and $\Gamma_7$ of opposite parity. 
The group $\mathbf{D}_{4h}$ is generated by the rotations $C_{4z}$ and $C_{2y}$, and by inversion $I$. The corep matrices have the form\cite{BC-book,Lax-book,Sam19-PRB} 
\begin{equation}
\label{corep-matrices}
  \left.\begin{array}{l}
        \hat{\cal D}_{\Gamma_6}(C_{4z})=\hat D^{(1/2)}(C_{4z}),\quad \hat{\cal D}_{\Gamma_6}(C_{2y})=\hat D^{(1/2)}(C_{2y}),\medskip\\ 
        \hat{\cal D}_{\Gamma_7}(C_{4z})=-\hat D^{(1/2)}(C_{4z}),\quad \hat{\cal D}_{\Gamma_7}(C_{2y})=\hat D^{(1/2)}(C_{2y}),
        \end{array}\right.
\end{equation}
where $\hat D^{(1/2)}(R)=e^{-i\theta(\bm{n}\hat{\bm{\sigma}})/2}$ is the spin-1/2 representation of a counterclockwise rotation $R$ through an angle $\theta$ about an axis $\bm{n}$. 
Note that $\hat{\cal D}_{\Gamma_7}$ is not equivalent to $\hat D^{(1/2)}$, reflecting the fact that $\Gamma_7$ is a non-pseudospin corep.

We obtain from Eqs. (\ref{phi-transform-g-1D}), (\ref{ab-parities}), and (\ref{corep-matrices}) that the singlet and triplet interband components transform independently from each other, are odd in $\bk$, 
and satisfy the following equations:
\begin{equation}
\label{ab-equations}
  \left.\begin{array}{l}
  \tilde\alpha(\bk)=-\tilde\alpha(C_{4z}^{-1}\bk),\quad \tilde\alpha(\bk)=\tilde\alpha(C_{2y}^{-1}\bk),\\ \\
  \tilde{\bm{\beta}}(\bk)=-C_{4z}\tilde{\bm{\beta}}(C_{4z}^{-1}\bk),\quad \tilde{\bm{\beta}}(\bk)=C_{2y}\tilde{\bm{\beta}}(C_{2y}^{-1}\bk).
  \end{array}\right.
\end{equation}
Here we used the identity $\hat{D}^{(1/2),\dagger}(R)\hat{\sigma}_i\hat{D}^{(1/2)}(R)=\sum_{j=1}^3R_{ij}\hat{\sigma}_j$, where $\hat R$ is the $3\times 3$ orthogonal rotation matrix. 
The simplest, polynomial in $\bk$, solution of the equations (\ref{ab-equations}) can be easily found:
$$
  \tilde\alpha(\bk)\propto k_xk_yk_z,\quad \tilde{\bm{\beta}}(\bk)\propto[k_x,-k_y,a(k_x^2-k_y^2)k_z],
$$
where $a$ is a real constant. Similarly, one can obtain all other expressions in Tables I and II.

\section{Gap nodes in the two-band case}
\label{sec: Gap nodes}

How does the unusual structure of the interband pairing found in the previous section affect the quasiparticle energy gap? The excitation spectrum is obtained by diagonalizing the Bogoliubov-de Gennes (BdG) Hamiltonian obtained
from Eq. (\ref{H-mean-field-general}):
$$
  \hat H_{BdG}=\left(\begin{array}{ccc}
                          \hat{\cal H}_{11} & \cdots & \hat{\cal H}_{1N} \\
                          \vdots & \ddots & \vdots \\
                          \hat{\cal H}_{N1} & \cdots & \hat{\cal H}_{NN} 
                          \end{array}\right),
$$
where $N$ is the number of superconducting bands,
$$
  \hat{\cal H}_{nn'}(\bk)=\left(\begin{array}{cc}
                                \hat\xi_n(\bk)\delta_{nn'} & \hat\Delta_{nn'}(\bk) \\
                                \hat\Delta^\dagger_{n'n}(\bk) & -\hat\xi_n(\bk)\delta_{nn'}
                                \end{array}\right),
$$
and $\hat\xi_n(\bk)=\xi_n(\bk)\hat\sigma_0$. There is a gap node at the wave vector $\bk$ if $\det\hat H_{BdG}(\bk)=0$. 

In the two-band case, the BdG Hamiltonian is an $8\times 8$ matrix in the tensor product of the band, particle-hole, and conjugacy spaces:
\begin{equation}
\label{H_BdG}
  \hat H_{BdG}=\left(\begin{array}{cccc}
                          \hat\xi_1 & \hat\Delta_{11} & 0 & \hat\Delta_{12} \\
                          \hat\Delta_{11}^\dagger & -\hat\xi_1 & \hat\Delta_{21}^\dagger & 0 \\
                          0 & \hat\Delta_{21} & \hat\xi_2 & \hat\Delta_{22} \\
                          \hat\Delta_{12}^\dagger & 0 & \hat\Delta_{22}^\dagger & -\hat\xi_2
                     \end{array}\right),
\end{equation}
with the gap functions given by Eq. (\ref{two-band-gaps}). The $2\times 2$ conjugacy blocks in this matrix all commute with each other, which greatly simplifies the calculation. We find 
$\det\hat H_{BdG}(\bk)=\det\hat{\cal R}(\bk)$, where 
\begin{eqnarray*}
  \hat{\cal R} &=& \hat\xi_1^2\hat\xi_2^2+\hat\xi_1^2\hat\Delta_{22}^\dagger\hat\Delta_{22}+\hat\xi_2^2\hat\Delta_{11}^\dagger\hat\Delta_{11}\\
    && +\hat\xi_1\hat\xi_2(\hat\Delta_{12}^\dagger\hat\Delta_{12}+\hat\Delta_{21}^\dagger\hat\Delta_{21})\\
    && +(\hat\Delta_{11}^\dagger\hat\Delta_{22}^\dagger-\hat\Delta_{12}^\dagger\hat\Delta_{21}^\dagger)(\hat\Delta_{11}\hat\Delta_{22}-\hat\Delta_{12}\hat\Delta_{21}).
\end{eqnarray*}
It is easy to show that all terms in $\hat{\cal R}$ are proportional to the unit matrix: $\hat{\cal R}(\bk)=R(\bk)\hat\sigma_0$, where
\begin{eqnarray*}
  R(\bk) &=& \xi_1^2\xi_2^2+\xi_1^2|\psi_2|^2+\xi_2^2|\psi_1|^2+2\xi_1\xi_2|\tilde\Delta|^2 \\
    && +|\psi_1\psi_2-\tilde\Delta^2|^2,
\end{eqnarray*}
$\psi_1(\bk)=\eta_1\alpha_1(\bk)$, $\psi_2(\bk)=\eta_2\alpha_2(\bk)$, and 
$$
  \tilde\Delta(\bk)=\tilde\eta\sqrt{\tilde\alpha^2(\bk)+\tilde{\bm{\beta}}^2(\bk)}
$$ 
is the measure of the interband pairing strength. Therefore, $\det\hat H_{BdG}(\bk)=R^2(\bk)$.
Choosing the interband order parameter to be real positive and putting $\eta_{n}=|\eta_{n}|e^{i\varphi_{n}}$ and $\alpha_n(\bk)=|\alpha_n(\bk)|e^{i\zeta_n(\bk)}$ (note that $\zeta_n=0$ or $\pi$, since $\alpha_n$ are real), 
we finally obtain: 
\begin{equation}
\label{BDG-det-final}
  \det\hat H_{BdG}=(r_1^2+r_2^2+r_3^2)^2,
\end{equation}
where
\begin{eqnarray}
\label{r_123}
  && r_1=\xi_1\xi_2-|\psi_1\psi_2|+|\tilde\Delta|^2,\nonumber\\
  && r_2=\xi_1|\psi_2|+\xi_2|\psi_1|,\\
  && r_3=\sqrt{2|\psi_1\psi_2||\tilde\Delta|^2(1-\cos\Phi)}\quad\nonumber
\end{eqnarray}
are real functions and $\Phi=\varphi_1+\varphi_2+\zeta_1+\zeta_2$. 

The BdG Hamiltonian (\ref{H_BdG}) has a zero eigenvalue if 
\begin{equation}
\label{node-condition}
  r_1(\bk)=r_2(\bk)=r_3(\bk)=0.
\end{equation}
In three-dimensional momentum space, this can generically happen only at isolated points, corresponding to point gap nodes. However, in our case the conditions (\ref{node-condition}) can be satisfied along certain lines, 
due to the special structure of $r_{1,2,3}$, see the examples below. In other systems with interband pairing and TR symmetry breaking, the gap can vanish on a whole surface in the momentum space 
(``Bogoliubov Fermi surface'').\cite{Bogoliubov-FS} 

To achieve analytical progress without losing much generality, we focus on the case of the $(\Gamma_6^\pm,\Gamma_6^\pm)$ or $(\Gamma_7^\pm,\Gamma_7^\pm)$ bands with
a quasi-2D dispersion: $\xi_n(\bk)=(k_x^2+k_y^2-k_{F,n}^2)/2m$, $k_{F,1}<k_{F,2}$. We also set $a=0$ in the intraband and interband basis functions and, 
in order to account for the lattice periodicity, replace $k_z$ by $\sin(k_zd)$, where $d$ is the lattice period along the $z$ axis, so that $\tilde{\bm{\beta}}^2\propto\sin^2(k_zd)$ in both the $s$- and $d$-wave cases. 
The calculation details can be found in the Appendix.

\begin{figure}
\includegraphics[width=7.5cm]{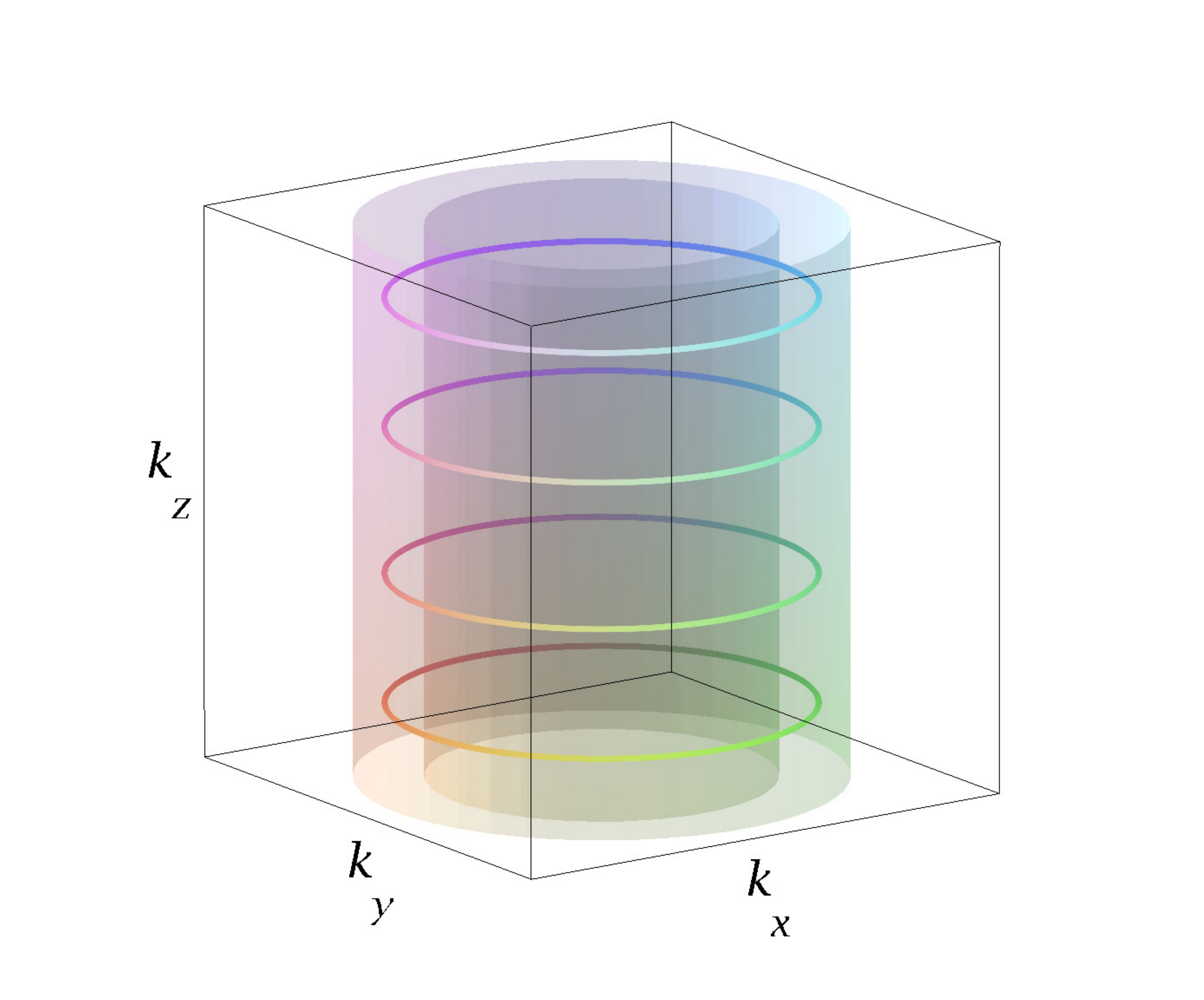}
\caption{(Color online) Lines of nodes in a two-band $s$-wave superconductor, for a strong interband pairing. The circular cylinders are the Fermi surfaces in the two bands.}
\label{fig: s-wave}
\end{figure}

For the $s$-wave pairing, we assume that the superconducting state is TR invariant, with $\varphi_1=\varphi_2=0$ or $\pi$ (our argument actually works more generally for $\varphi_1=-\varphi_2$, 
which is the case for all stable states of the three-band Ginzburg-Landau model in Ref. \onlinecite{ST10}), and that $\zeta_1=\zeta_2=0$ (no accidental zeros of $\alpha_1$ and $\alpha_2$), therefore $r_3=0$. 
The remaining equations $r_1=0$ and $r_2=0$ have solutions if the interband pairing is sufficiently strong:
$$
  |\tilde\Delta|^2>|\psi_1\psi_2|.
$$
These solutions correspond to four horizontal circular lines of nodes located between the two cylindrical Fermi surfaces, see Fig. \ref{fig: s-wave}.

\begin{figure}
\includegraphics[width=7.5cm]{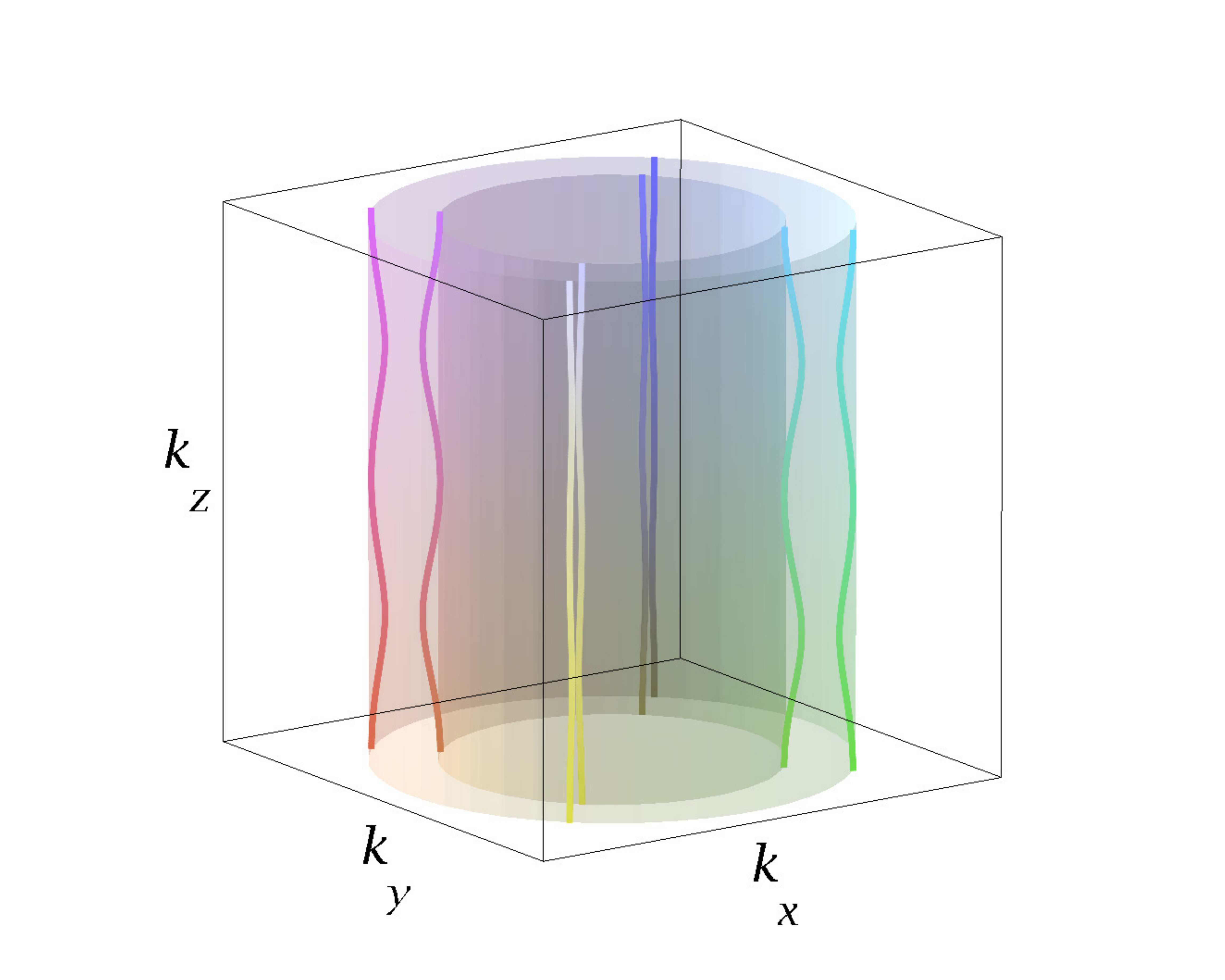}
\caption{(Color online) Lines of nodes in a two-band $d_{x^2-y^2}$-wave superconductor, for a weak interband pairing.}
\label{fig: d-wave-1}
\end{figure}

\begin{figure}
\includegraphics[width=7.5cm]{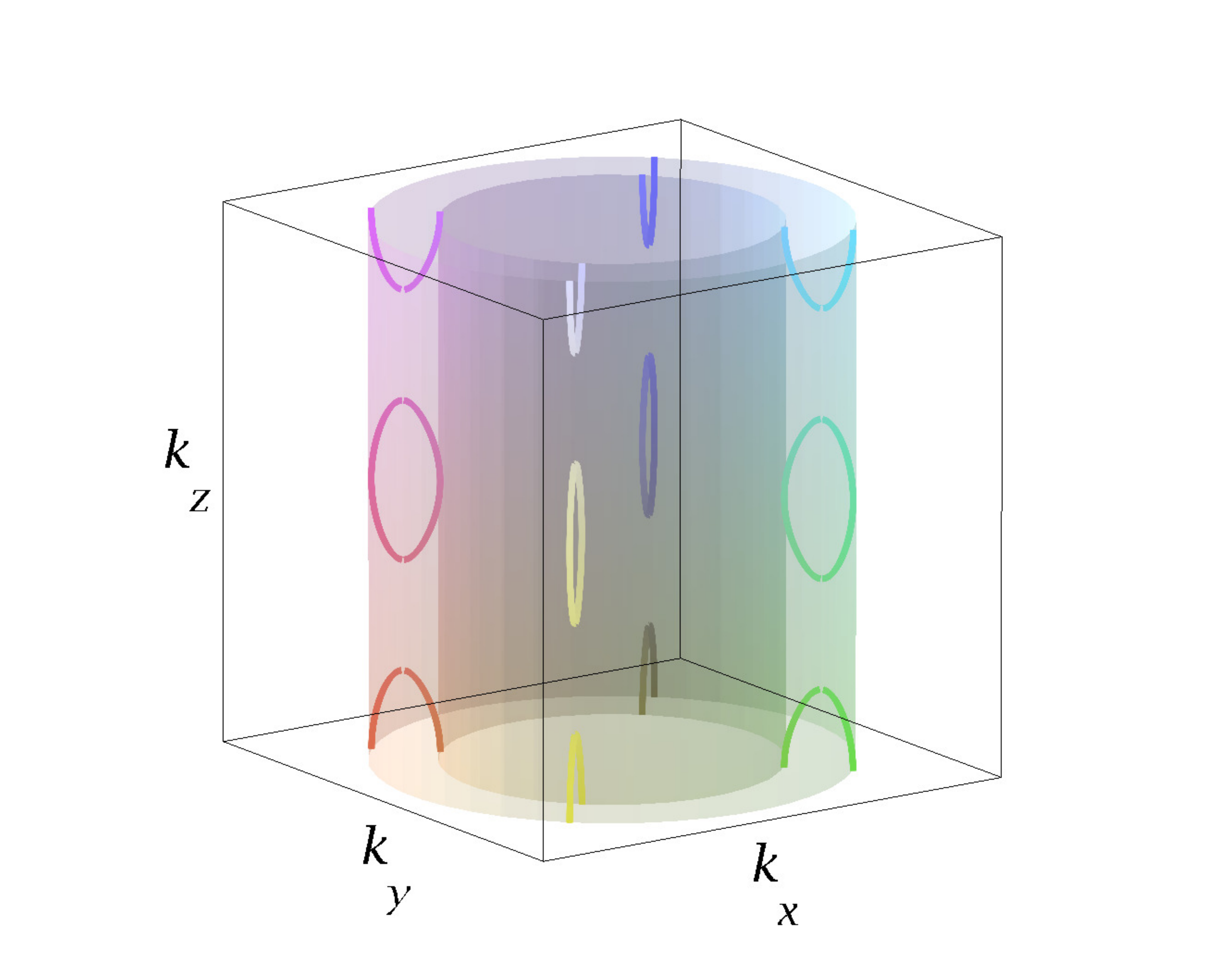}
\caption{(Color online) Reconnected lines of nodes in a two-band $d_{x^2-y^2}$-wave superconductor, for a strong interband pairing.}
\label{fig: d-wave-2}
\end{figure}

For the $d$-wave pairing, the intraband gaps $\psi_1$ and $\psi_2$ vanish at $|k_x|=|k_y|$ for the symmetry reasons, therefore $r_2=r_3=0$ in the diagonal planes. 
The remaining equation $r_1=0$ takes the form 
$$
  \xi_1\xi_2=-|\tilde\Delta|^2.
$$
If the interband pairing is weak, then the lines of nodes are deformed away from the Fermi surfaces into the ``interband space'', 
see Fig. \ref{fig: d-wave-1}. However, at a sufficiently strong interband pairing a topological transition takes place: the lines of nodes originating from the two bands touch and then reconnect in a different configuration, 
forming vertical nodal loops, as shown in Fig. \ref{fig: d-wave-2}. 

The lines of nodes result in a linear behaviour of the quasiparticle density of states at low energies, $N(E)\propto E$. That in turn produces characteristic power laws in
the temperature dependence of thermodynamic and transport properties. For instance, for the electronic specific heat one has $C(T)\propto T^2$, at $T\to 0$ (Refs. \onlinecite{SU-review} and \onlinecite{TheBook}). 
There might be some novel features associated with the nodal line reconnection transition in the $d$-wave case, but those are beyond the scope of the present study.

\section{Conclusions}
\label{sec: Conclusions}

To summarize, we showed how to construct the Bloch bases in crystals with spin-orbit coupling, for twofold degenerate electron bands which do not transform under the point group operations like the pure spin-$1/2$ states.
The consequences of the ``non-pseudospin'' character of the bands for superconductivity include such remarkable features as nonzero interband triplet components for $s$-wave and $d$-wave pairing, 
odd-parity singlet and even-parity triplet interband gaps, etc. A sufficiently strong interband pairing profoundly affects the nodal structure, changing the line node topology
in the $d$-wave case and producing lines of nodes in the $s$-wave case, which can be seen in the temperature dependence of thermodynamic and transport properties in the superconducting state. 

\acknowledgments
This work was supported by a Discovery Grant 2015-06656 from the Natural Sciences and Engineering Research Council of Canada.

\appendix*

\section{Lines of nodes}

We consider the $(\Gamma_6^\pm,\Gamma_6^\pm)$ or $(\Gamma_7^\pm,\Gamma_7^\pm)$ bands with 
$$
  \xi_n(\bk)=\frac{k_\perp^2-k_{F,n}^2}{2m},\quad k_\perp=\sqrt{k_x^2+k_y^2},
$$ 
$k_{F,2}=\varrho k_{F,1}$, and $\varrho>1$. Using different pairs of bands and/or taking into account the band modulation along $k_z$ will not change the results qualitatively.

\subsection{$s$-wave pairing}

Neglecting the possibility of accidental zeros of the intraband gap functions, one can set $\zeta_1=\zeta_2=0$ at all $\bk$. In order for $r_3$ to vanish, we require that $\cos(\varphi_1+\varphi_2)=1$, which is 
satisfied, in particular, for the TR invariant superconducting states with $\varphi_1=\varphi_2=0$ or $\pi$. Assuming that $\xi_1(\bk)>\xi_2(\bk)$, the equations $r_1=0$ and $r_2=0$ take the form 
\begin{equation}
\label{s-equations}
  \left.\begin{array}{l}
         \xi_1=|\psi_1|\sqrt{\dfrac{|\tilde\Delta|^2}{|\psi_1\psi_2|}-1}, \\
         \xi_2=-|\psi_2|\sqrt{\dfrac{|\tilde\Delta|^2}{|\psi_1\psi_2|}-1}.
         \end{array}\right.
\end{equation}
If $|\tilde\Delta|^2>|\psi_1\psi_2|$, then the two surfaces defined by these equations may intersect along a line, or lines, in the momentum space. These lines are located between the two Fermi surfaces, where $\xi_1>0$ 
and $\xi_2<0$.

To obtain an explicit solution of Eq. (\ref{s-equations}), we assume that the intraband gap functions are constant, with $\alpha_n(\bk)=1$ and the same gap magnitudes $|\eta_1|=|\eta_2|=\eta$ in both bands, 
whereas for the interband gap functions 
one can put, according to Table I, $\tilde\alpha(\bk)=1$ and $\tilde{\bm{\beta}}(\bk)=b[\sin\theta\sin(k_zd),-\cos\theta\sin(k_zd),0]$, where $\theta=\arctan(k_y/k_x)$ and $b$ is a real constant. 
Therefore, 
\begin{eqnarray*}
  && \psi_1(\bk)=\eta e^{i\varphi},\quad \psi_2(\bk)=\eta e^{-i\varphi},\\
  && \tilde\Delta(\bk)=\tilde\eta\sqrt{1+b^2\sin^2(k_zd)}.
\end{eqnarray*}
We substituted $\sin(k_zd)$ for $k_z$ in order to account for the lattice periodicity. 

It is straightforward to show that the equations (\ref{s-equations}) have the following four solutions: 
\begin{equation}
\label{s-lines}
  \left.\begin{array}{l}
         k_z=\pm\dfrac{1}{d}\arcsin\sigma,\ \pm\dfrac{1}{d}(\pi-\arcsin\sigma),\\
         k_\perp=k_{F,1}\sqrt{\dfrac{\varrho^2+1}{2}}, 
        \end{array}\right.
\end{equation}
where
$$
  \sigma=\frac{1}{\delta b}\sqrt{\left(\frac{\varrho^2-1}{2\epsilon}\right)^2-\delta^2+1},\quad \epsilon=\frac{2m\eta}{k_{F,1}^2},\quad \delta=\frac{\tilde\eta}{\eta}. 
$$
These solutions correspond to four circular lines of nodes shown in Fig. 1 and exist if the parameters of the system are such that $\sigma<1$.

\subsection{$d_{x^2-y^2}$-wave pairing}

For the symmetry reasons, $\psi_1=\psi_2=0$ in the diagonal planes $|k_x|=|k_y|$, therefore $r_2=r_3=0$ in these planes, 
regardless of the values of the order parameter phases. We are left with just one equation $r_1=0$, which takes the form
\begin{equation}
\label{d-equation}
  \xi_1\xi_2=-|\tilde\Delta|^2.
\end{equation}
The interband gap functions can be chosen as follows: $\tilde\alpha(\bk)=\cos(2\theta)$ and $\tilde{\bm{\beta}}(\bk)=b[\sin\theta\sin(k_zd),\cos\theta\sin(k_zd),0]$, 
see Table II, therefore $\tilde\Delta(\bk)=\tilde\eta b|\sin(k_zd)|$ in the diagonal planes. 

The equation (\ref{d-equation}) has the following solutions:
\begin{equation}
\label{d-solutions}
  \left.\begin{array}{l}
         |k_x|=|k_y|=q_{+}(k_z)k_{F,1},\medskip \\
         |k_x|=|k_y|=q_{-}(k_z)k_{F,1},
        \end{array}\right.
\end{equation}
where
$$
  q_\pm(k_z)=\frac{1}{2}\left[\varrho^2+1\pm\sqrt{(\varrho^2-1)^2-4\tilde\epsilon^2b^2\sin^2(k_zd)}\right]^{1/2},
$$
and $\tilde\epsilon=2m\tilde\eta/k_{F,1}^2$. In the absence of the interband pairing, i.e., at $\tilde\epsilon=0$, the expressions (\ref{d-solutions}) describe four pairs of vertical lines of nodes on the two Fermi surfaces.
With increasing $\tilde\epsilon$, these lines get deformed and partially leave the Fermi surfaces, as shown in Fig. 2. At $\tilde\epsilon=\tilde\epsilon_c=(\varrho^2-1)/2b$, 
the lines of nodes touch, at $k_z=\pm\pi/2d$. Finally, at $\tilde\epsilon>\tilde\epsilon_c$, the lines of nodes reconnect into nodal loops (Fig. 3), which shrink, but never completely disappear, 
as the interband pairing strength increases.


\begin{thebibliography}{99}

\bibitem{VG85}
G. E. Volovik and L. P. Gor'kov, Zh. Eksp. Teor. Fiz. \textbf{88}, 1412 (1985) [Sov. Phys. JETP \textbf{61}, 843 (1985)].

\bibitem{SU-review}
M. Sigrist and K. Ueda, Rev. Mod. Phys. \textbf{63}, 239 (1991).

\bibitem{TheBook}
V. P. Mineev and K. V. Samokhin, \textit{Introduction to Unconventional Superconductivity} (Gordon and Breach, London, 1999).

\bibitem{And84}
P. W. Anderson, Phys. Rev. B \textbf{30}, 4000 (1984).

\bibitem{UR85}
K. Ueda and T. M. Rice, Phys. Rev. B \textbf{31}, 7114 (1985).

\bibitem{Kittel-book}
C. Kittel, \textit{Quantum Theory of Solids} (Wiley, 1987).

\bibitem{multiorbital-SC}
Y. Wan and Q.-H. Wang, Europhys. Lett. \textbf{85}, 57007 (2009); M. H. Fischer, New J. Phys. \textbf{15}, 073006 (2013); A. Ramires and M. Sigrist, Phys. Rev B \textbf{94}, 104501 (2016); 
T. Nomoto, K. Hattori, and H. Ikeda, Phys. Rev. B \textbf{94}, 174513 (2016); W. Huang, Y. Zhou, and H. Yao, Phys. Rev. B \textbf{100}, 134506 (2019).

\bibitem{j-3-2-pairing}
P. M. R. Brydon, L. Wang, M. Weinert, and D. F. Agterberg, Phys. Rev. Lett. \textbf{116}, 177001 (2016);
H. Kim \textit{et al.}, Sci.  Adv. \textbf{4}, eaao4513 (2018).

\bibitem{Sam19-PRB}
K. V. Samokhin, Phys. Rev. B \textbf{100}, 054501 (2019).

\bibitem{BlountSam}
E. I. Blount, Phys. Rev. B \textbf{32}, 2935 (1985); K. V. Samokhin, Ann. Phys. (N.Y.) \textbf{385}, 563 (2017). 

\bibitem{BC-book}
C. J. Bradley and A. P. Cracknell, \textit{The Mathematical Theory of Symmetry in Solids} (Oxford University Press, Oxford, 2010).

\bibitem{unusual-ASOC}
M. Smidman, M. B. Salamon, H. Q. Yuan, and D. F. Agterberg, Rep. Prog. Phys. \textbf{80}, 036501 (2017);
K. V. Samokhin, Ann. Phys. (N. Y.) \textbf{407}, 179 (2019).

\bibitem{irreps}
We use the ``chemical'' notation for the single-valued irreps corresponding to the pairing channels, reserving the $\Gamma$ notation for the double-valued coreps describing the symmetry of the Bloch bands.

\bibitem{ST10}
V. Stanev and Z. Te\v{s}anovi\'c, Phys. Rev. B \textbf{81}, 134522 (2010).

\bibitem{Lax-book}
M. Lax, \textit{Symmetry Principles in Solid State and Molecular Physics} (Dover Publications, New York, 2001).

\bibitem{Bogoliubov-FS}
G. E. Volovik, Phys. Lett. A \textbf{142}, 282 (1989); P. M. R. Brydon, D. F. Agterberg, H. Menke, and C. Timm, Phys. Rev B \textbf{98}, 224509 (2018).

\end{thebibliography}
\end{document}